\documentclass[prc,showpacs,showkeys,superscriptaddress,floatfix]{revtex4}
\usepackage{graphicx,amssymb,amsmath}

\begin{document}
\title{Relationship between X(5)-models and the interacting boson model}

\author{Jos\'e Barea} \email{jose.barea@iem.cfmac.csic.es}
\affiliation{Instituto de Estructura de la Materia, CSIC, Serrano 123,
E-28006 Madrid, Spain}
\affiliation{Departamento de F\'{\i}sica At\'omica, Molecular y
  Nuclear, Facultad de F\'{\i}sica, Universidad de Sevilla,
  Apartado~1065, 41080 Sevilla, Spain}

\author{Jos\'e M. Arias} \email{ariasc@us.es}
\affiliation{Departamento de F\'{\i}sica At\'omica, Molecular y
  Nuclear, Facultad de F\'{\i}sica, Universidad de Sevilla,
  Apartado~1065, 41080 Sevilla, Spain}

\author{Jos\'e Enrique Garc\'{\i}a-Ramos}
\email{enrique.ramos@dfaie.uhu.es} \affiliation{Departamento de
  F\'{\i}sica Aplicada, Universidad de Huelva, 21071 Huelva, Spain}

\begin{abstract}
  The connections between the X(5)-models (the original X(5) using
  an infinite square well, X(5)-$\beta^8$,  X(5)-$\beta^6$,
  X(5)-$\beta^4$, and 
   X(5)-$\beta^2$), based on particular solutions of the geometrical
  Bohr Hamiltonian with harmonic potential in the $\gamma$ degree of
  freedom, and the
  interacting boson model (IBM) are explored. This work is the natural
  extension of the work presented in \cite{Garc08} for the E(5)-models.
  For that purpose, a quite
  general one- and two-body IBM Hamiltonian 
  is used
  and a numerical fit to the different X(5)-models energies is
  performed, later on the obtained wave functions are used to calculate
  B(E2) transition rates. It is shown that within the IBM one can
  reproduce well the results for energies and B(E2) transition
  rates 
  obtained with all these X(5)-models, although the agreement
  is not so impressive as for the E(5)-models.  
  From the fitted IBM parameters the corresponding energy surface can
  be extracted and it is obtained that, surprisingly, only the X(5)
  case corresponds in the moderate large N limit to an energy surface
  very close to
  the one expected for a critical point, while the rest of models seat
  a little farther.
\end{abstract}
\maketitle

\section{Introduction}
\label{sec-intro}
In recent years the connection between the Bohr-Mottelson
(BM) collective model 
\cite{Bohr52,Bohr53,Bohr69} and the interacting boson model (IBM)
\cite{Arim76,Arim78,Scho78,Iach87} has been the subject of many studies
\cite{Diep80a,Diep80b,Gino80a,Gino80b,Kirs82,Gino82,Rowe05,
Lope96,Aria07,Rowe04b,Rowe04c,Rowe05b,TR09}. The BM collective model is
built on the assumption that the 
nucleus is composed by a set of strongly interacting fermions that can
be treated as a quantum liquid. The surface of such a liquid is
characterized in terms of the Hill-Wheeler shape variables $(\beta,\gamma)$
\cite{Hill53} and the Euler angles. Under this approximation, nuclear
excitations are small amplitude vibrations and rotations,
corrected by the coupling between them \cite{Eise70}. The IBM was
designed to describe the collective quadrupole degrees of
freedom in medium mass and heavy nuclei. The IBM
Hamiltonian was written from the beginning in second quantization
form in terms of the
generators of the U(6) algebra, subtended by $s$ and $d$ bosons which
carry angular momenta $0$ and $2$, respectively \cite{Iach87}. Therefore, the
connection between both models is not evident. An approximated
connection comes from considering the IBM as the second quantization
of the shape variables $(\beta,\gamma)$ \cite{Jans74}. During the
eighties many studies on the connection between both models were
done. The intrinsic state
formalism \cite{Diep80a,Diep80b,Gino80a,Gino80b,Kirs82,Gino82} was
used, but also the complete set of eigenstates \cite{Mosh80} was analyzed through
a Holstein-Primakoff transformation \cite{Klei81,Klei82} or by a isometric
transformation \cite{Asse83}. More
recently, the problem of the mapping between both models has been addressed by
Rowe and collaborators \cite{Rowe05}. Both models presents three special
limits that can be solved easily and for which the connection between the
models is known.  These three cases are: i) the BM
anharmonic vibrator and the dynamical symmetry U(5) IBM limit, ii)
the BM $\gamma$-unstable deformed rotor and the dynamical O(6) IBM
limit, and iii) the BM axial rotor and the dynamical symmetry O(6)
IBM limit including $Q\cdot Q\cdot Q$ interactions
\cite{Isac99,Rowe05,TR09}. Note that, although traditionally accepted,
the correspondence of the dynamical symmetry SU(3) IBM limit to a
submodel of the BM has never been explicitly probed
\cite{Rowe05}.  Each of these limits is assigned to a particular shape
using the Hill-Wheeler variables $(\beta,\gamma)$ \cite{Hill53}:
spherical, deformed with $\gamma$-instability, and axially deformed,
respectively.  For transitional situations the correspondence between
the two models is difficult and a possible way to establish a mapping
between BM and IBM is through numerical studies.   

Among the transitional Hamiltonians, a specially interesting case
occurs when it describes a critical point in the transition from a
given shape to another. In general, for such a situation, where the
structure of the system can change abruptly by applying a small
perturbation, both, the BM and the IBM, have to be solved numerically.
However, few years ago Iachello proposed schematic Bohr Hamiltonians
that intend to describe different critical points and that can be
solved exactly in terms of the zeroes of Bessel functions. The first of
these models is known as E(5) \cite{Iach00}. E(5) is designed to
describe the critical point at the transition from spherical to
deformed $\gamma$-unstable shapes. The potential to be used in the
differential Bohr equation is assumed to be $\gamma-$independent and,
for the $\beta$ degree of freedom an infinite square well is taken.
Similar models were proposed later on by Iachello,
called X(5) and Y(5) \cite{Iach01,Iach03}, to describe the
critical points between spherical and axially deformed shapes and
between axial and triaxial deformed shapes, respectively. All these
models give rise to spectra and electromagnetic transition rates that
depend on a couple of free parameters (including a scale). In spite of 
their simplicity, some experimental examples were found 
\cite{Cast00,Cast01}, just after the appearance of these models.

In this work, we concentrate on X(5) and related models (the
connection between E(5) and related
models and the IBM was already deeply studied in \cite{Garc08}). 
The formulation of X(5) attracted immediately
attention both experimentally and theoretically. Soon after the
introduction of the X(5) model, the nucleus $^{152}$Sm was proposed
by Casten and Zamfir \cite{Cast01} as a realization of it. Other
experimental examples proposed are: 
$^{150}$Nd, $^{152-154}$Gd, $^{130}$Ce, $^{162}$Yb, $^{166}$Hf, 
$^{178}$Os, $^{226}$Ra, and $^{226}$Th \cite{Cast07,Cejn10}, although
  the last two candidates could be better described combining quadrupole and
  octupole degrees of freedom \cite{Bizz04,Bizz08,Bona05}.
Concerning
theoretical extensions of X(5), Bonatsos and collaborators studied 
a sequence of potentials of the type $\beta^{2n}$,
that allows to go from the vibrational limit,
$n=1$, to X(5), $n \rightarrow \infty$ \cite{Bona04b}. In particular, in
Ref.~\cite{Bona04b} spectra and transition rates for the potentials  
of the type $\beta^{2n}$, with $n\ge 1$, 
$\beta^2, \beta^4, \beta^6$ and, $\beta^8$, are given explicitly and compared
with the original X(5) (infinite square well potential) case. 
Other extension of X(5) is X(3), which is a rigid version of X(5)
\cite{Bona06}. In reference \cite{Mccu05} the authors studied the
connection between X(5) and a two parameters free IBM calculation. 
In reference \cite{Mccu06} the authors compared
X(5)-$\beta^2$, X(5)-$\beta^4$, and X(3) also with a restricted two parameter
IBM calculation with a number of bosons $N=10$.
In references \cite{Bona07a,Bona07b} the authors treat an exactly separable
version of the Davidson potential and of the X(5) potential,
respectively.  Finally, in reference \cite{Capr05} the author studied
the effect that the $\beta-\gamma$ coupling has in solving the BM equation
for the X(5) potential.  As
mentioned above, all these models are produced in the BM scheme and a
natural question is to ask for the correspondence of them with the
IBM. Is the IBM able to produce the same spectra and transition
rates? If yes, does the obtained IBM Hamiltonian correspond to a critical
point? This work is intended to answer these questions for the X(5)
and related models ($-\beta^8$, $-\beta^6$, $-\beta^4$ and, 
$-\beta^2$ potentials)
and analyze the convergence as a function of the boson number, $N$.

For that purpose, a large set of X(5) and related models results for
excitation energies and transition rates are taken as reference for
numerical fits of the general IBM  Hamiltonian.
This procedure will allow to establish the IBM Hamiltonian which best
fit the different X(5)-$\beta^{2n}$ models and their relation with the
critical points.

The paper is organized as follows: in section \ref{sec-fit} the
fitting procedure is described and the obtained results are commented.
Section \ref{sec-crit} is devoted to study the energy surfaces of the
fitted IBM Hamiltonians and to analyze them in relation to the
critical point. Finally, in section \ref{sec-conclu} the summary and
conclusions of this work are presented.

\begin{figure}[hbt]
  \centering
  \includegraphics[width=10cm]{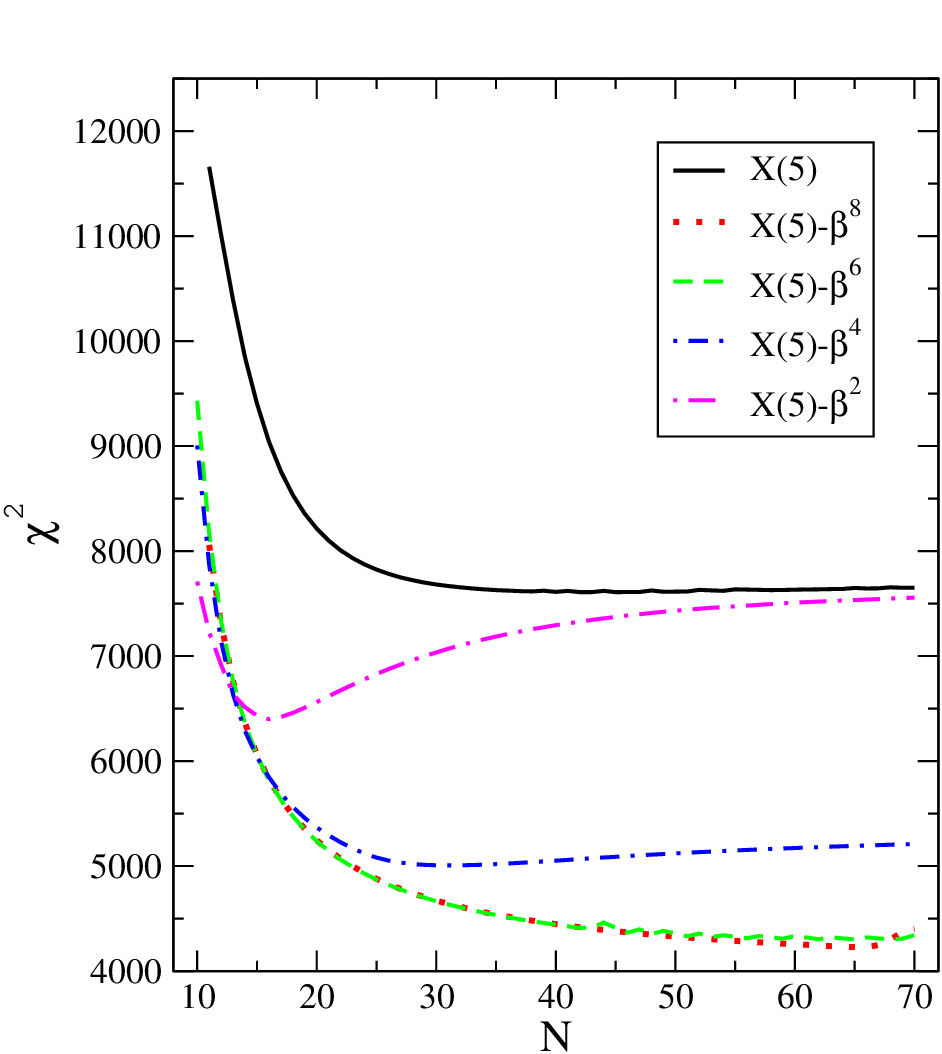}
  \caption{(Color online). $\chi^2$ for the IBM fit to the energy
    levels of the different X(5)-models, as a function of $N$.}
  \label{fig-chi2-x5-ener}
\end{figure}

\section{The IBM fit to X(5)-models}
\label{sec-fit}
\subsection{The model}
\label{sec-model}

The most general, including up to two-body terms, IBM Hamiltonian can
be written in multipolar form as,

\begin{eqnarray}
  \label{ham1}
  \hat H&=&\varepsilon_d \hat n_d +
  \kappa_0 \hat P^\dag \hat P
  +\kappa_1 \hat L\cdot \hat L+
  \kappa_2 \hat Q \cdot \hat Q+
  \kappa_3 \hat T_3\cdot\hat T_3 +\kappa_4 \hat T_4\cdot\hat T_4
\end{eqnarray}  
where $\hat n_d$ is the $d$ boson number operator, and
\begin{eqnarray}
  \label{P}
  \hat P^\dag&=&\frac{1}{2} ~ (d^\dag \cdot d^\dag - s^\dag \cdot s^\dag), \\
  \label{L}
  \hat L&=&\sqrt{10}(d^\dag\times\tilde{d})^{(1)},\\
  \label{Q}
  \hat Q&=& (s^{\dagger}\times\tilde d
  +d^\dagger\times\tilde s)^{(2)}-
  \frac{\sqrt{7}}{2}(d^\dagger\times\tilde d)^{(2)},\\
  \label{t3}
  \hat T_3&=&(d^\dag\times\tilde{d})^{(3)},\\
  \label{t4}
  \hat T_4&=&(d^\dag\times\tilde{d})^{(4)}.
\end{eqnarray} 
The symbol $\cdot$ stands for the scalar product, defined as $\hat
T_L\cdot \hat T_L=\sum_M (-1)^M \hat T_{LM}\hat T_{L-M}$ where $\hat
T_{LM}$ is the $M$ component of the operator $\hat T_{L}$.
The operator $\tilde\gamma_{\ell m}=(-1)^{m}\gamma_{\ell -m}$ (where
$\gamma$ refers to $s$ and $d$ bosons) is introduced to ensure the
correct tensorial character under spatial rotations.

\begin{figure}[hbt]
  \centering
  \includegraphics[width=14cm]{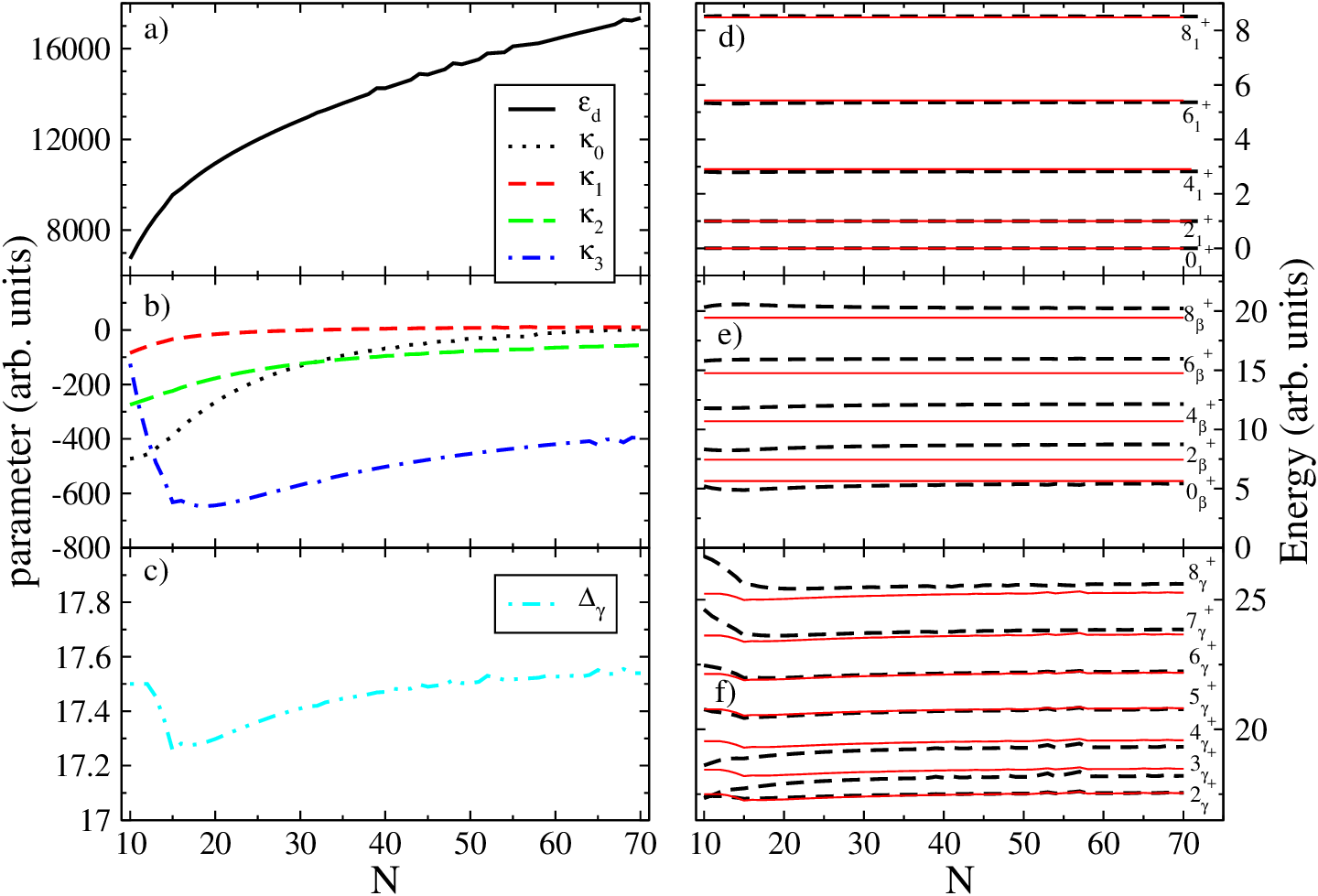}
  \caption{(Color online). Parameters (left panels (a), (b) and (c))
  and excitation energies (right panels) of the
  ground (d), beta (e), and gamma (f) bands for the  X(5) case, as a
  function of $N$. In the right panels, the continuous red
  lines are the X(5) results, while the dashed black lines are the
  fitted results.}
  \label{fig-par-ener-x5}
\end{figure}

The electromagnetic transitions can also be analyzed in the framework
of the IBM. In particular, in this work we will focus on the E2
transitions. The most general E2 transition operator including up to
one body terms is written as,
\begin{equation}
  \label{te2}
  \hat T^{E2}_M=e_{eff}\left[(s^\dag \times 
    \tilde{d}+d^\dag\times\tilde{s})^{(2)}_M+
    \chi(d^\dag\times\tilde{d})^{(2)}_M\right],
\end{equation}
where $e_{eff}$ is the boson effective charge and $\chi$ is a
structure parameter. In this work $\chi$ will be kept fixed to
the SU(3) value $-\sqrt{7}/2$. 

%%%R1
Although one could use the most general IBM Hamiltonian (\ref{ham1})
to describe the X(5)-models  a
natural question is whether it is possible or not to reduce the number of free
parameters. A priori, it is not obvious which terms of the Hamiltonian
can be taken out. To answer this question one can rewrite the Hamiltonian
(\ref{ham1}) in terms of Casimir operators (the 
definitions for the Casimir operators have been taken from
\cite{Fran94}):
\begin{eqnarray}
  \nonumber
  \hat H&=& \frac{\kappa_0}{4} N(N+4)+
  \big (\varepsilon_d +\frac{18}{35} \kappa_4 \big)\,\hat C_1[U(5)]+
  \frac{18}{35} \kappa_4 \,\hat C_2[U(5)]\\
  &+&\big(\kappa_1-\frac{3}{8}\kappa_2-\frac{\kappa_3}{10}-\frac{\kappa_4}{14}\big)\,\hat
  C_2[O(3)] +\frac{\kappa_2}{2}\,\hat C_2[SU(3)]
+ \big(\frac{\kappa_3}{2}-\frac{3}{14} \kappa_4\big)\,\hat C_2[O(5)]
  -\frac{\kappa_0}{4}\,\hat C_2[O(6)].
  \label{ham-cas}
\end{eqnarray}
It is expected that for the description of the X(5)-models one will
need the contribution of the three IBM dynamical symmetry chains. In
the Hamiltonian (\ref{ham-cas}) two contributions come
from the U(5) algebra, the linear and the quadratic Casimir
operators. Therefore, it becomes a reasonable ``antsatz'' to remove the
U(5) quadratic Casimir operator. That implies to fix
$\kappa_4=0$. Note that its contribution to the rest of Casimir
operators is absorbed by the rest of the parameters.

\begin{figure}[hbt]
  \centering
  \includegraphics[width=14cm]{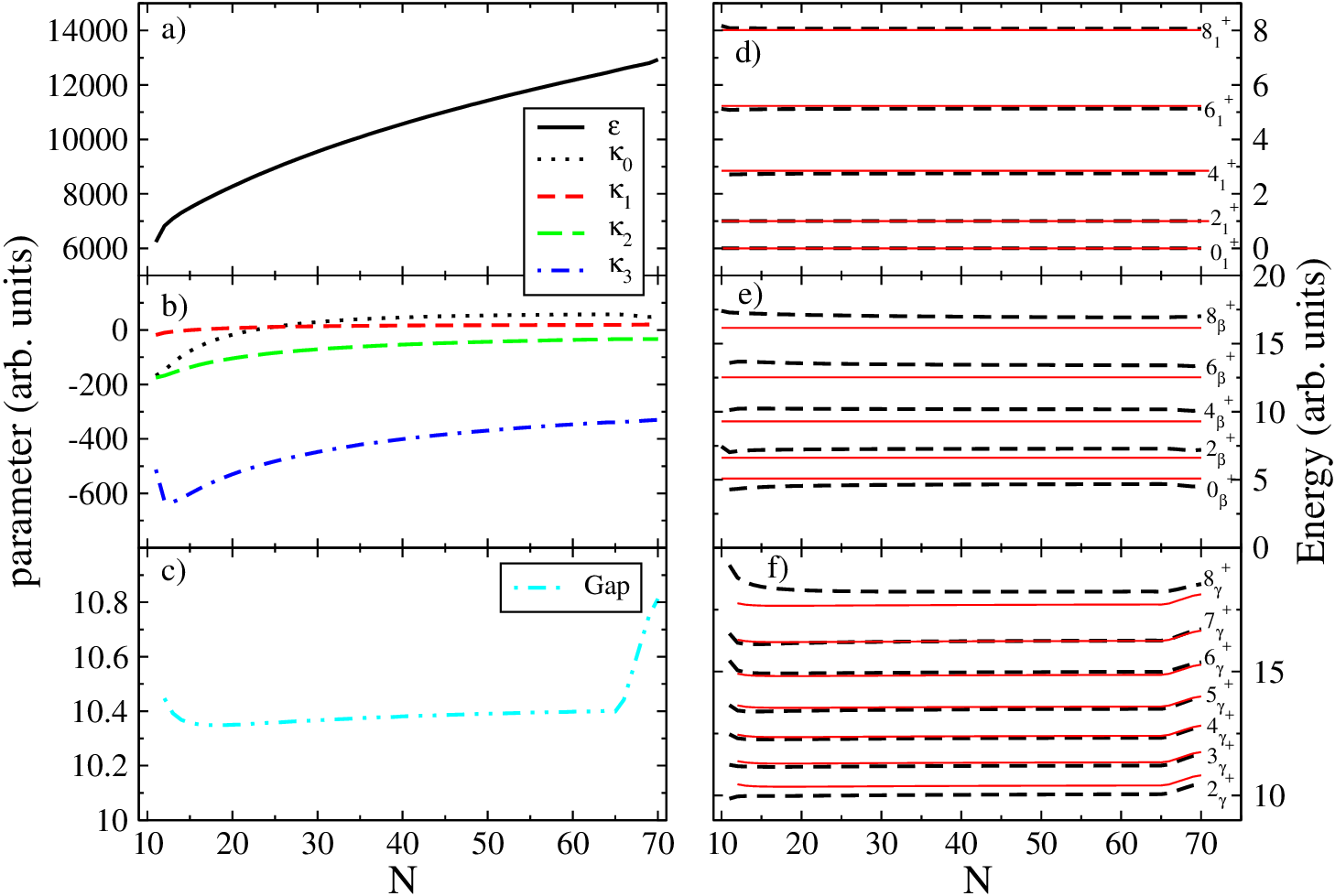}
  \caption{(Color online). Same as Fig.~\ref{fig-par-ener-x5} but for the
  X(5)-$\beta^8$ case.}
  \label{fig-par-ener-x5-b8}
\end{figure}

\subsection{The fitting procedure}
In this section we describe the procedure for getting the IBM
Hamiltonian parameters that best fit the different X(5)-models.

The $\chi^2$ test is used to perform the fitting. The $\chi^2$
function is defined in the standard way,
\begin{equation}
  \label{chi2}
  \chi^2=\frac{1}{N_{data}-N_{par}}\sum_{i=1}^{N_{data}}\frac{(X_i
    (data)-X_i (IBM))^2}{\sigma_i^2},
\end{equation} 
where $N_{data}$ is the number of data, from a specific X(5)-model,
to be fitted, $N_{par}$ is 
the number of parameters used in the IBM fit, $X_i(data)$
is an energy level (or a B(E2) value) taken from a particular
X(5)-model, $X_i(IBM)$ is the corresponding calculated IBM value,
and $\sigma_i$ is an arbitrary error assigned to each $X_i(data)$.

At this point it is necessary to explain how do we treat the $\gamma$
band head in the X(5) model. The position of this band is not determined
by the model, therefore an extra parameter should be introduced for
determining the gamma band head, specifically the location of the band head
$2_\gamma^+$ state. Consequently, 
besides the IBM Hamiltonian parameters, there is this extra free
parameter, $\Delta_\gamma$, in the fit. This parameter gives the
optimum position of the $\gamma$ band in a given X(5)-model for which
the best IBM fit is obtained.

In order to perform the fit, we minimize the $\chi^2$ function for the
energies, using
$\varepsilon_d$, $\kappa_0$, $\kappa_1$, $\kappa_2$, and $\kappa_3$, 
as free parameters of the IBM Hamiltonian and
$\Delta_\gamma$ as free parameter for the position of the gamma band
head. 
Note that besides the qualitative argument presented in section
\ref{sec-model} to justify 
the election of $\kappa_4=0$, we have extensively explored other
possibilities, as for example $\kappa_4\neq 0$ \cite{Unpu10}. For this
election the improvement in $\chi^2$ is not worthy and for some
particular cases even affects negatively. In addition, in this
case the  $\chi^2$ function is very flat and the correlation that exists
between the free parameters generates non physical oscillations
in the value of the fitted parameters. We have also explored the case
$\kappa_3=0$ and $\kappa_4=0$, which produces a important increase in
the $\chi^2$ value. The $\kappa_4=0$ selection produces a smooth and
consistent behavior
of the fitted parameters, as can be observed in figures
\ref{fig-par-ener-x5}, \ref{fig-par-ener-x5-b8},
\ref{fig-par-ener-x5-b6}, \ref{fig-par-ener-x5-b4}, and  
\ref{fig-par-ener-x5-b2}.

For doing the fit and the minimization of the $\chi^2$ function the
MINUIT \cite{minuit} code has been used. It allows to
minimize any multi-variable function. 

The labels for the energy levels follow the usual notation introduced
for the X(5) model \cite{Iach00}: $s$ enumerates the zeroes of the $\beta$ part
of the wave function, and $n_\gamma$ enumerates the number of $\gamma$
phonons. The set of levels included in the fit for the different
X(5)-models are:
\begin{itemize}
\item For the ground state band, $s=1, n_\gamma=0$, all the states
  with angular momentum smaller 
  than $10$. An arbitrary $\sigma= 0.001$ is used for
  these states except for the $2_1^+$ state for which $\sigma=0.0001$
  is used. This latter value allows to normalize all the IBM energies
  to $E(2_1^+)=1$. Note that the energy of the state $2_1^+$ is fixed
  arbitrarily to $1$ (remind that the spectrum is calculated up to a
  global scale factor).

\item For the beta band, $s=2, n_\gamma=0$, all the states with
  angular momentum smaller 
  than $10$. An arbitrary $\sigma=0.01$ is used for these
  states.

\item For the $s=3, n_\gamma=0$ band, which can be identified with
  the $\beta\beta$ band, all the states with angular momentum smaller
  than $10$. An arbitrary $\sigma=0.01$ is used for these
  states.

\item For the gamma band, $s=1, n_\gamma=1$, all the states with
  angular momentum smaller
  than $9$. An arbitrary $\sigma=0.01$ is used for these
  states. 
\end{itemize}
With this selection, the number of energy levels included into the fit,
$N_{data}$, is equal to $21$. Note that the state $0_1^+$ is not an
actual data to be reproduced because we are interested just in
excitation energies and therefore the ground state is naturally fixed
to zero in both, X(5)-models and IBM. In Table \ref{tab-energ-fit}
the states included in the fit are explicitly given.

\begin{table}
  \begin{tabular}{|c|c|c|}
    \hline
    Band    &Error          & States\\
    \hline
    $s=1, n_\gamma=0$ &$\sigma=0.0001$& $2_1^+$ \\
    &$\sigma=0.001$ & $0_1^+, 4_1^+, 6_1^+, 8_1^+ $ \\
    \hline
    $s=2, n_\gamma=0$ &$\sigma=0.01$ & $0_2^+, 2_2^+, 4_2^+, 6_2^+, 8_2^+ $ \\
    \hline
    $s=3, n_\gamma=0$ &$\sigma=0.01$ & $0_3^+, 2_{3,4}^+, 4_4^+, 6_4^+, 8_4^+ $ \\
    \hline
    $s=1, n_\gamma=1$ &$\sigma=0.01$ & $2_{4,3}^+, 3_1^+, 4_3^+, 5_1^+,
    6_3^+, 7_1^+, 8_3^+ $ \\
    \hline
  \end{tabular}
  \caption{States included in the energy fit. In the states labeled
    with two subindexes, the first one corresponds to X(5), whereas
    the second to the rest of models.}
  \label{tab-energ-fit}
\end{table}
The ordering index for the even L states in the $\gamma$, $\beta$, and
$\beta\beta$ bands is unknown a priori, because of the undetermined position
of the X(5)-$\gamma$ bands. However, our best fit always provides a 
$\gamma$ band above the $\beta$ band, but below the $\beta\beta$ band,
that generates the same ordering independently of the particular
X(5)-model and number of bosons. The only exception happens for X(5),
where the $2^+$ of the $\gamma$ band is above the $2^+$ state in
the $\beta\beta$ band, {\it i.e.}~they correspond to $2^+_4$ and
$2^+_3$ respectively.

Once the IBM Hamiltonian is fixed for each X(5)-model by fitting the
energy levels, the $\chi^2$ function for the B(E2) values is
calculated without any additional fitting. The two parameters in the
E2 operator (\ref{te2}) are $e_{eff}$, which is fixed
to give $B(E2; 2_1^+\rightarrow 0_1^+)=100$, and  $\chi$, which is
fixed to the SU(3) value $-\sqrt{7}/2$.  The
computed transitions are enlisted in Table \ref{tab-be2-fit}. Note
that only transitions between states with $n_\gamma=0$ states are
considered.

\begin{table}
\begin{tabular}{|c|c|c||c|c|c|}
  \hline
  & $s_i$ &$s_f$ & &$s_i$ &$s_f$ \\ 
  \hline
  $B(E2:2_1^+\rightarrow  0_1^+)$& 1 & 1 & $B(E2:2_2^+\rightarrow  0_1^+)$& 2 & 1 \\                  
  $B(E2:4_1^+\rightarrow  2_1^+)$& 1 & 1 & $B(E2:2_2^+\rightarrow  0_2^+)$& 2 & 2 \\                  
  $B(E2:6_1^+\rightarrow  4_1^+)$& 1 & 1 & $B(E2:4_2^+\rightarrow  2_1^+)$& 2 & 1 \\                     
  $B(E2:8_1^+\rightarrow  6_1^+)$& 1 & 1 & $B(E2:4_2^+\rightarrow  4_1^+)$& 2 & 1 \\                     
  $B(E2:0_2^+\rightarrow  2_1^+)$& 2 & 1 & $B(E2:4_2^+\rightarrow  2_2^+)$& 2 & 2 \\                   
  $B(E2:2_2^+\rightarrow  2_1^+)$& 2 & 1 & & & \\                   
  \hline
\end{tabular}
\caption{B(E2) transitions to be calculated. $n_\gamma=0$ in all
  the cases.}
\label{tab-be2-fit}
\end{table}

\subsection{The results}

We have performed fits of the IBM Hamiltonian (\ref{ham1}) parameters plus
$\Delta_\gamma$ for different values of the number of bosons, $N$, so
as to reproduce as well as possible the energies of the states given
in Table \ref{tab-energ-fit}. These states are generated by the different
X(5)-models: X(5), X(5)-$\beta^8$, X(5)-$\beta^6$, X(5)-$\beta^4$, 
and X(5)-$\beta^2$.

As mentioned before, $\varepsilon_d$, $\kappa_0$, $\kappa_1$,
$\kappa_2$,  $\kappa_3$, and  $\Delta_\gamma$ are free
parameters in a $\chi^2$ fit to the energy
levels produced by the different X(5)-models ($\kappa_4$ was fixed
to zero as discussed in the preceding section). In figure
\ref{fig-chi2-x5-ener} the value of the $\chi^2$ for the best fit to
the different X(5)-models as a function of $N$ is shown. The different
lines in figure \ref{fig-chi2-x5-ener} correspond to the fit to
different X(5)-models as stated in the legend box. 
It is clearly
observed that for any $N$ the best agreement is obtained for the
X(5)-$\beta^6$ and X(5)-$\beta^8$  cases, which present a $\chi^2$
function almost identical for any value of N. The $\chi^2$ function increases for
X(5)-$\beta^4$ and X(5)-$\beta^2$, up to reach X(5) which has the higher
$\chi^2$ value and therefore the worst degree of agreement. 
Anyway, the differences between the different X(5)-models are
smaller than a factor $2$ in $\chi^2$. 
It is worth noting that these results change slowly with the boson
number and in all cases the $\chi^2$ function saturates to a given
value in the large N limit.
The situation presented here is somehow different to the analysis of
the $E(5)$-models \cite{Garc08} where there is a
monotonous behavior in the $\chi^2$ values in  passing from E(5)-$\beta^4$,
which presents the lowest value, to E(5)-$\beta^6$, E(5)-$\beta^8$,
and E(5), where the maximum appear.

\begin{figure}[hbt]
  \centering
  \includegraphics[width=14cm]{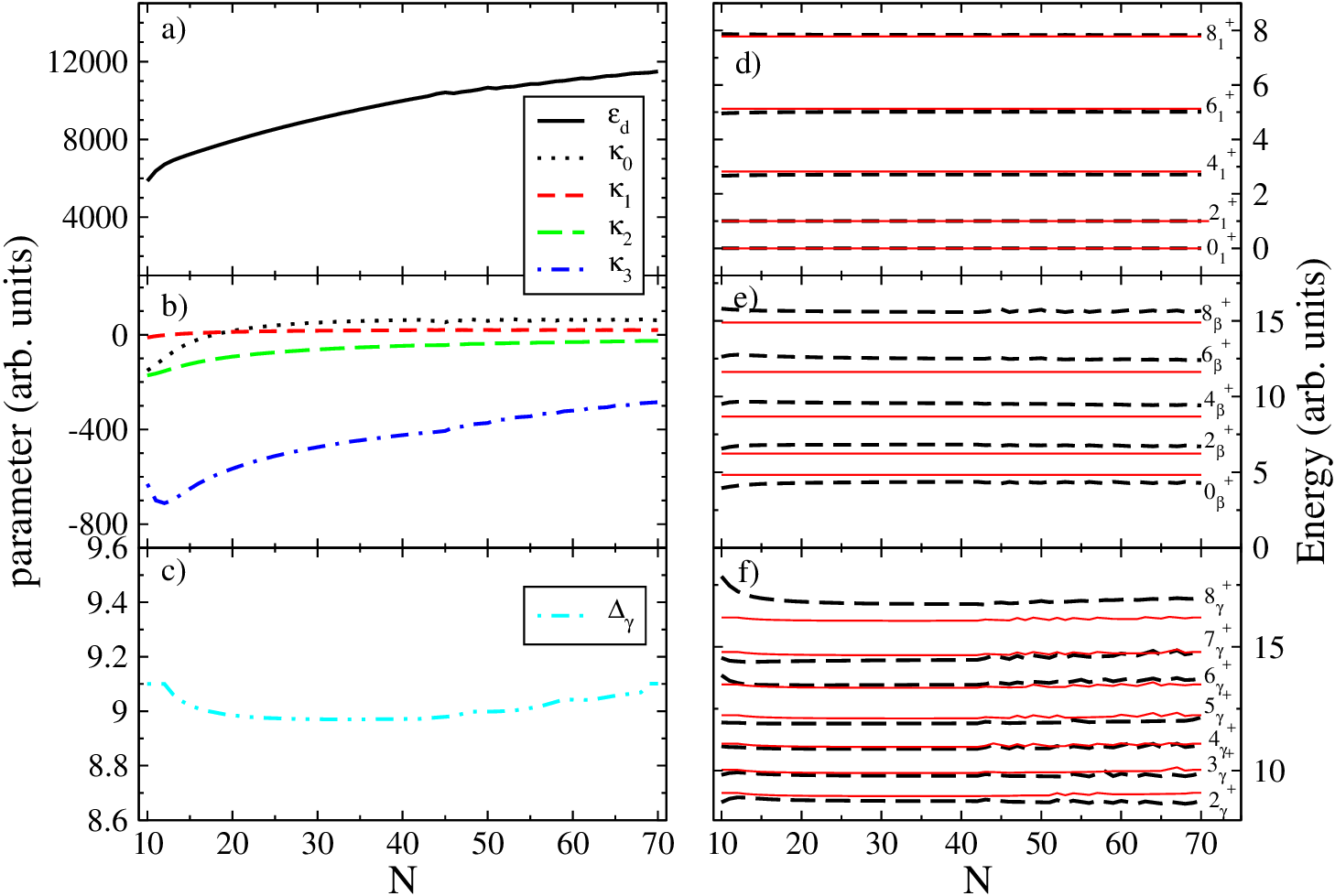}
  \caption{(Color online). Same as Fig.~\ref{fig-par-ener-x5} but for the
  X(5)-$\beta^6$ case.}
  \label{fig-par-ener-x5-b6}
\end{figure}

\begin{figure}[hbt]
  \centering
  \includegraphics[width=14cm]{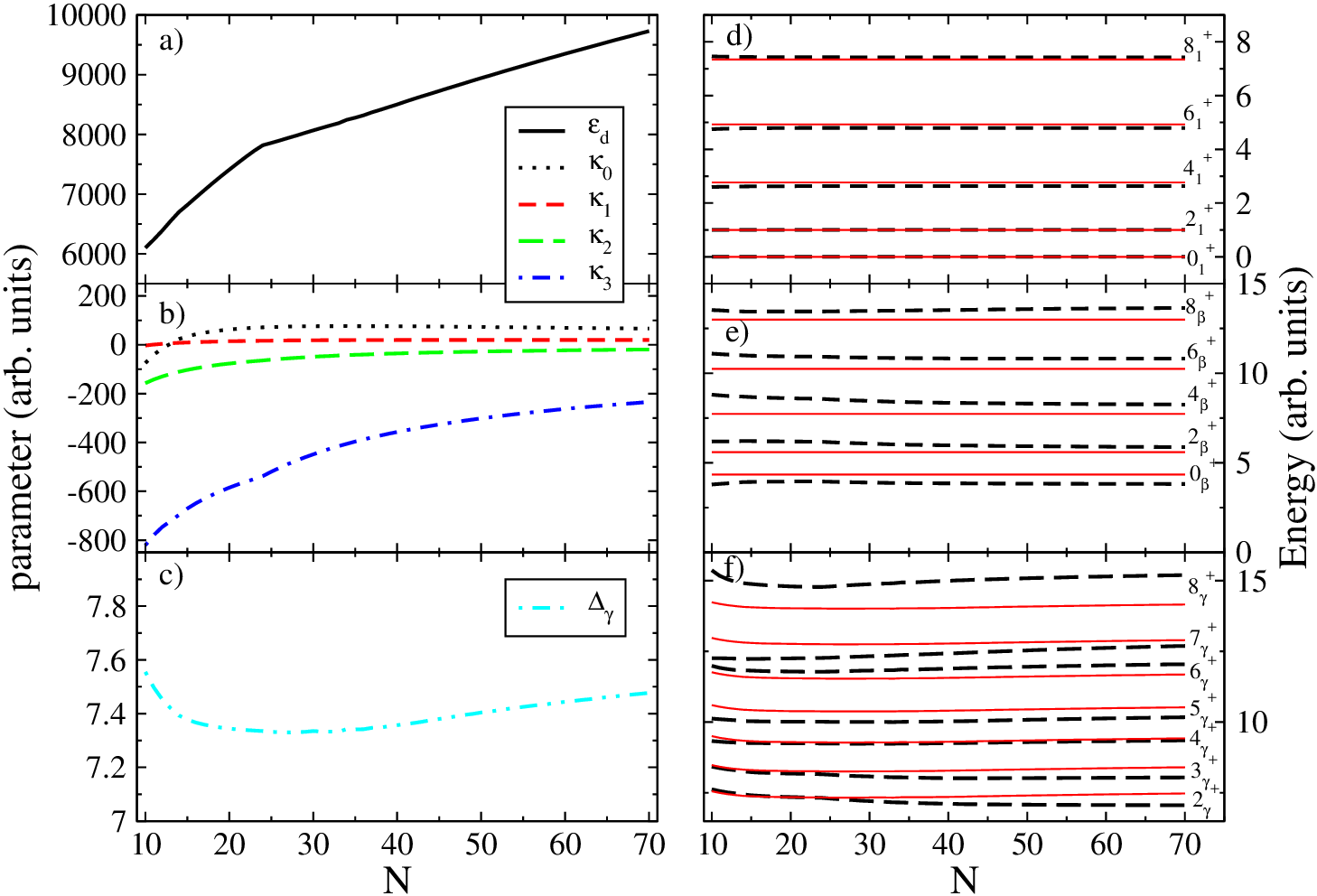}
  \caption{(Color online). Same as Fig.~\ref{fig-par-ener-x5} but for the
  X(5)-$\beta^4$ case.}
  \label{fig-par-ener-x5-b4}
\end{figure}

\begin{figure}[hbt]
  \centering
  \includegraphics[width=14cm]{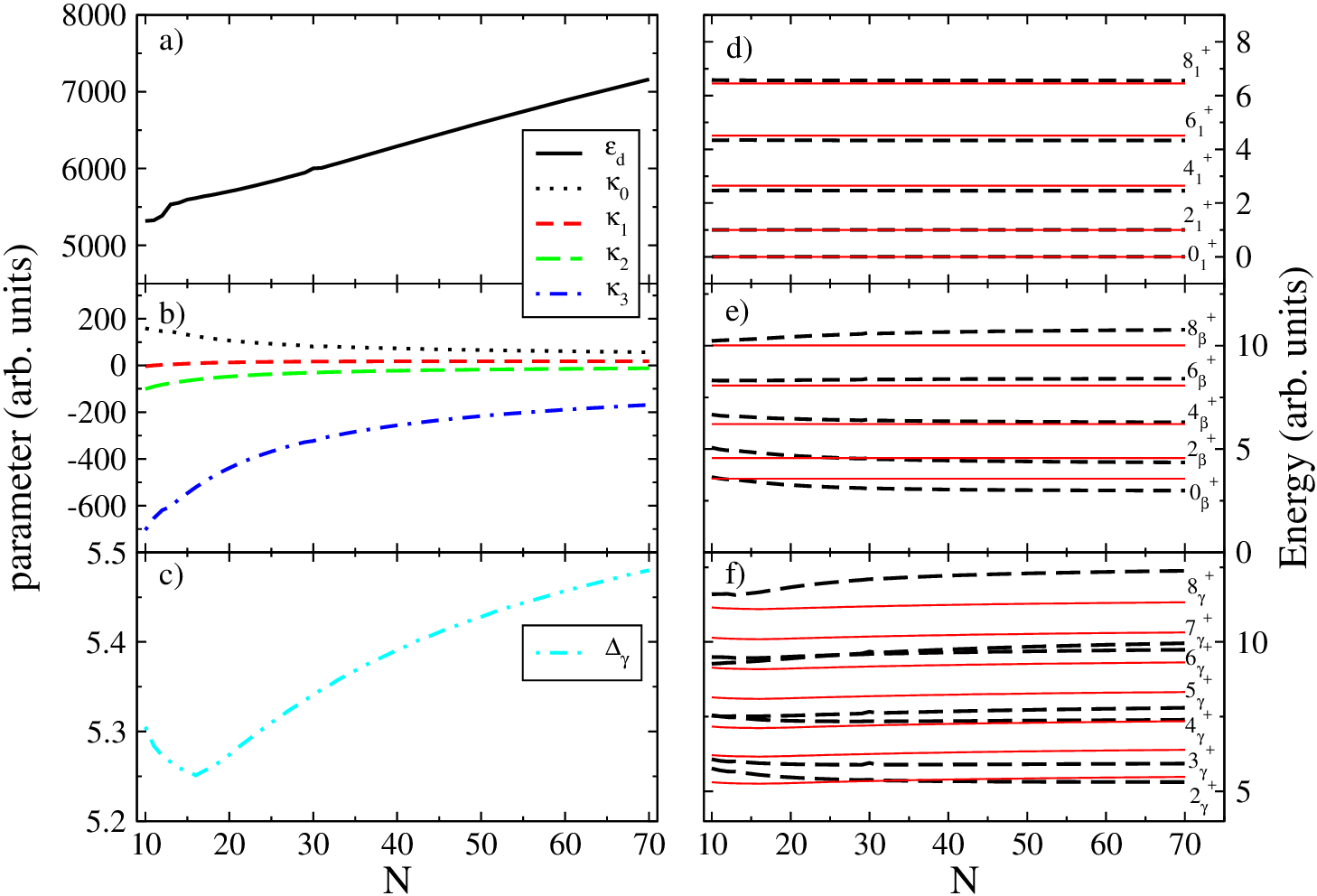}
  \caption{(Color online). Same as Fig.~\ref{fig-par-ener-x5} but for the
  X(5)-$\beta^2$ case.}
  \label{fig-par-ener-x5-b2}
\end{figure}

In figures \ref{fig-par-ener-x5}, \ref{fig-par-ener-x5-b8},
\ref{fig-par-ener-x5-b6}, \ref{fig-par-ener-x5-b4}, and
\ref{fig-par-ener-x5-b2} the Hamiltonian parameters (panels
(a), (b) and (c)) and the
excitation energies for the ground (panel (d)), beta (panel (e)) and
gamma (panel (f)) bands are plotted for the cases
of X(5), X(5)-$\beta^8$, X(5)-$\beta^6$, X(5)-$\beta^4$, and
X(5)-$\beta^2$, respectively.
The Hamiltonian parameters present clear analogies in its
behavior in all the X(5)-models.  $\varepsilon_d$ increases
continuously on the whole range of 
N for all the X(5)-models. $\kappa_1$, on the contrary, presents a
relatively small and modest value. $\kappa_0$ and $\kappa_2$ show a
very smooth variation till reaching a value of saturation, in all the
cases the tendency is to increase, except in the case of  $\kappa_0$ for
X(5)-$\beta^2$ which tends to decrease. $\kappa_3$ in almost all the
cases tends to increase all the way. In the X(5) case,
Fig.~\ref{fig-par-ener-x5}, $\kappa_3$ shows up a lowering for small
values of N while shows the already commented increasing behavior for larger
N values. A similar behavior is also observed for
$\Delta_\gamma$. Concerning the energy
levels, it is remarkable that
the plotted energies  have almost constant values regardless the value
of $N$, except for the $\gamma$ band in the low N region where the
energies smoothly move to a saturation value. The agreement for the ground
state band is very good but for the $\beta$ and $\gamma$ band
the agreement is poorer. In the X(5)-$\beta^8$ case,
Fig.~\ref{fig-par-ener-x5-b8}, $\kappa_3$ also presents a decreasing behavior
for very low values of N, while from there on, it monotonously
increases.  $\Delta_\gamma$ presents an almost constant value, except
for the largest values of N, where a increase (look at the small
scale) is shown. 
Regarding the energies, they remain almost
constant in the full range of $N$, except for the $\gamma$ band,
where, the increase of the $\Delta_\gamma$ value generates, for
large values on N, a corresponding increasing. The agreement of the
energies in
the ground state band is perfect and reasonable in the $\beta$ 
and $\gamma$ bands.
The X(5)-$\beta^6$ case, Fig.~\ref{fig-par-ener-x5-b6}, is very
similar to  X(5)-$\beta^8$, although here $\Delta_\gamma$ presents a
rather flat behavior with a minimum around $N=40$. Once more, the agreement of the
energies in
the ground state band is perfect and reasonable in the $\beta$ 
and $\gamma$ bands. The X(5)-$\beta^4$ case shown in
Fig.~\ref{fig-par-ener-x5-b4} presents a monotonous increase of
$\kappa_3$ and a behavior for $\Delta_\gamma$ almost identical to
X(5)-$\beta^6$. Here also the agreement of the energies in the ground
state band is good while the description of the $\beta$ and
$\gamma$ bands is poorer. 
Finally, X(5)-$\beta^2$, Fig.~\ref{fig-par-ener-x5-b2}, also presents a monotonous increase of
$\kappa_3$ whereas $\kappa_0$  has a smooth decreasing behavior till a
saturation value. In $\Delta_\gamma$ the behavior of X(5) is recovered,
with a smooth decrease up to $N=15$ and an smooth increase from there on. The
agreement of the energies in the ground state band is very good and reasonable for the $\beta$ 
and $\gamma$ bands. 

In panels (e) of Figs.~\ref{fig-par-ener-x5}, \ref{fig-par-ener-x5-b8},
\ref{fig-par-ener-x5-b6}, \ref{fig-par-ener-x5-b4}, and
\ref{fig-par-ener-x5-b2} one can see that 
the moment of inertia of the $\beta$ band 
increases from X(5), where it reaches the smallest value, till X(5)-$\beta^2$
where it is maximum.  In panels (d) and (f) of these figures one can also
observe how the moment of inertia of ground and $\gamma$ bands is
roughly stable for all the models.
This tendency is correctly reproduced by the IBM fits. 

It is worth remarking that the agreement of the energies in the
$\beta$ and the $\gamma$ bands is not clearly deteriorated when
ascending up in the band. In particular, in the $\gamma$ band the best
agreement is obtained for angular momenta values 4, 5, 6. Similar conclusions
are obtained for the $\beta\beta$ band. This fact has a important
consequence for the value of $\chi^2$ and supposes that, contrary to
expected, the higher states included in the fitting procedure for the 
$\beta$, $\gamma$, and $\beta\beta$ bands produce
smaller contributions to $\chi^2$ than the low lying states of these
bands. 

To have a clearer idea of the values of the Hamiltonian parameters
obtained in the fits we
present in Table \ref{table-parameters-not4t4} the parameters of the
IBM Hamiltonians that best fit the different X(5)-models for
N=50. Two facts are apparent, first the parameters for
X(5)-$\beta^8$ and X(5)-$\beta^6$ are amazingly similar and, second
the parameters change almost monotonously when going from X(5) to
X(5)$-\beta^2$.

\begin{table}
\begin{tabular}{|c|c|c|c|c|c|c|}
\hline
&$\varepsilon_d$ &$\kappa_0$ &$\kappa_1$&$\kappa_2$&$\kappa_3$&$\Delta_\gamma$\\
\hline
X(5)        &15420 &-31.7  &  7.5 & -77.1 & -454.8 & 17.5 \\
X(5)$-\beta^8$&11426 & 53.6  & 17.8 & -43.3 & -369.2 & 10.4 \\ 
X(5)$-\beta^6$&10661 & 58.5  & 19.9 & -38.5 & -372.7 &  9.2 \\
X(5)$-\beta^4$& 8942 & 72.9  & 19.6 & -27.8 & -301.4 &  7.4 \\
X(5)$-\beta^2$& 6595 & 65.9  & 18.2 & -17.6 & -216.6 &  5.4 \\
    \hline
  \end{tabular}
  \caption{Parameters (in arbitrary units) 
of the IBM Hamiltonian plus the excitation
    energy for the $\gamma$
    band head, $\Delta_\gamma$,
    that best fit the different X(5)-models for N=50.} 
  \label{table-parameters-not4t4}
\end{table}

\begin{figure}[hbt]
  \centering
  \includegraphics[width=10cm]{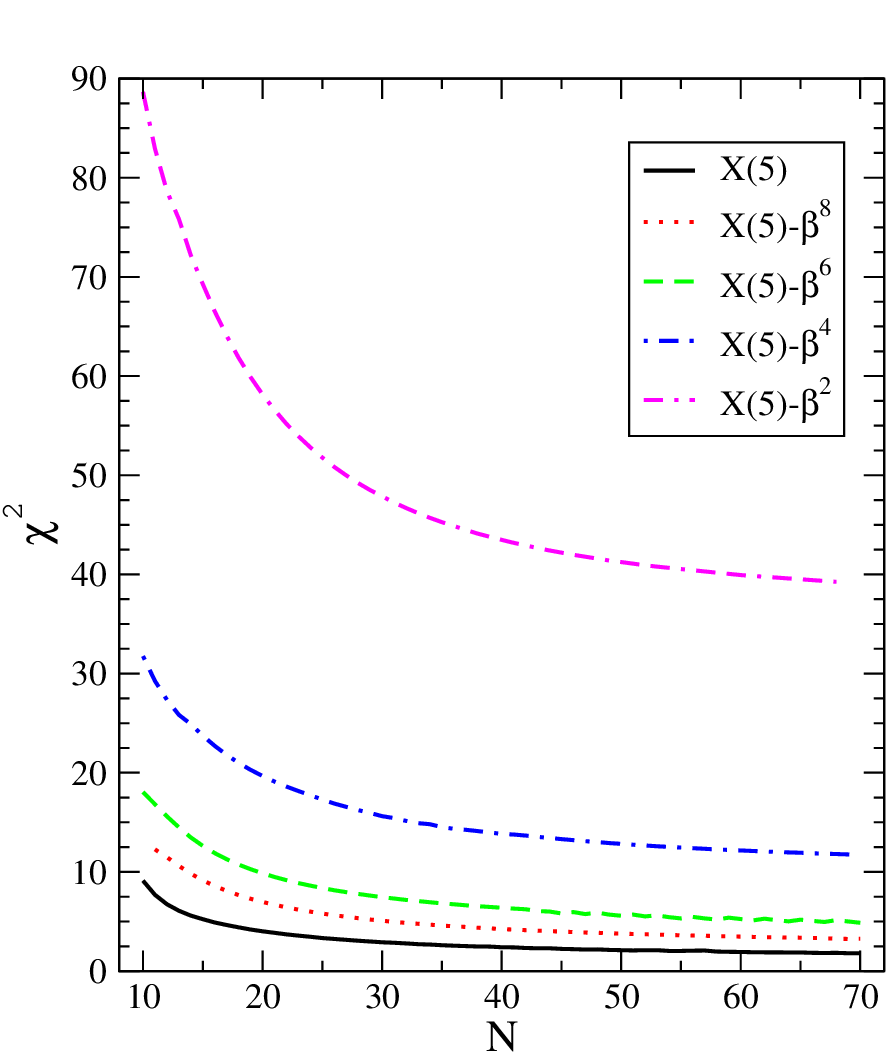}
  \caption{(Color online). $\chi^2$ values for the E2 transition
    rates for the different X(5)-models, as a function of $N$, calculated
    using the IBM electromagnetic transition operator $\hat
    T^{E2}_M=e_{\mbox{eff}} \big [(s^\dag
    \tilde d+ d^\dag\tilde
    s)-\sqrt{7}/2(d^\dag\times\tilde{d})^{(2)}_M$ \big ].}
  \label{fig-be2-chi2}
\end{figure}

As a test for the produced wave functions with the fitted IBM
Hamiltonian, they are used for calculating E2 transition
probabilities, B(E2). The effective charge in the
E2 operator (\ref{te2}) is fixed so as to give
$B(E2;2_1^+\rightarrow 0_1^+)=100$, thus no free parameters are left
in this calculation. For the B(E2)'s calculated (not a fit) a
$\chi^2$ value has been obtained for each X(5)-model with an
arbitrary $\sigma=10$.  In figure \ref{fig-be2-chi2} the
corresponding $\chi^2$ value is plotted as a function of $N$ for all
the X(5)-models considered.  Figure \ref{fig-be2-chi2}
shows a smooth dependence of $\chi^2$ on $N$. The $\chi^2$ value
decreases monotonically as $N$ increases for all the
X(5)-models. The best agreement is obtained for X(5) while the
worst is for X(5)-$\beta^2$.

For a quantitative comparison, the B(E2) values for selected
transitions with $N=50$ are shown in Table \ref{tab-be2-comp}. In this
table, it is clear the remarkable agreement between the IBM
calculations and those from the X(5) model. However as soon as we move to the
rest of models the agreement starts getting worse.
Some comments on these results are in order.
First, one observes that the B(E2)'s within the ground state band are
calculated qualitatively correct in all models. This agreement is
quantitatively good for the X(5) case and gets worse for the other
models. The X(5)-$\beta^2$ case is the worst, showing differences of
a factor of 2 for some transitions. The IBM seems to saturate to a too
small value as the angular momentum increases, therefore the deviation
increases with the angular momentum.   
The intraband transitions in the $\beta$ band exhibit the same kind
of agreement, where the larger discrepancy is for
$B(E2:4_2^+\rightarrow  2_2^+)$. The interband transitions once more
agree well qualitatively. In summary, one can say that the structure of the
wave functions is correctly captured by the IBM fit, specially for X(5), and
the calculations are able to reproduce the sequence of large/small
values (with few exceptions), confirming the appropriated structure of
the wave functions. 

\begin{table}[hbt]
  \begin{tabular}{|r||r|r||r|r||r|r||r|r||r|r|}
    \hline
    & X(5) & IBM & X(5)-$\beta^8$ &IBM  & X(5)-$\beta^6$ & IBM &
    X(5)-$\beta^4$ & IBM & X(5)-$\beta^2$ & IBM  \\
    \hline
$B(E2:2_1^+\rightarrow  0_1^+)$&100.0& 100.0&100.0& 100.0&100.0&100.0&100.0&100.0&100.0& 100.0\\
$B(E2:4_1^+\rightarrow  2_1^+)$&159.9& 155.4&163.4& 159.1&165.3&145.7&169.0&164.9&177.9& 174.6\\
$B(E2:6_1^+\rightarrow  4_1^+)$&198.2& 181.6&208.8& 189.2&214.6&160.8&226.2&199.9&255.2& 217.7\\
$B(E2:8_1^+\rightarrow  6_1^+)$&227.6& 199.1&247.3& 210.3&258.1&166.2&279.9&224.9&337.1& 249.2\\    
$B(E2:0_2^+\rightarrow  2_1^+)$& 62.4&  43.4& 74.7&  55.7& 81.0& 11.2& 93.2& 74.6&121.9& 106.7\\
$B(E2:2_2^+\rightarrow  2_1^+)$&  8.2&   4.4&  9.7&   7.3& 10.3&  0.9& 11.3& 15.4& 13.4&  33.4\\    
$B(E2:2_2^+\rightarrow  0_1^+)$&  2.1&   0.2&  2.2&   0.0&  2.2&  3.5&  2.0&  0.6&  1.6&   4.4\\    
$B(E2:2_2^+\rightarrow  0_2^+)$& 79.5&  56.1& 97.2&  67.5&106.0& 68.3&122.0& 69.3&155.7&  49.2\\    
$B(E2:4_2^+\rightarrow  2_1^+)$&  0.9&   0.2&  0.8&   0.0&  0.7&  5.2&  0.5&  0.7&  0.1&   2.1\\    
$B(E2:4_2^+\rightarrow  4_1^+)$&  6.1&   3.4&  7.7&   6.8&  8.4&  0.7&  9.6& 16.6& 12.4&  27.7\\   
$B(E2:4_2^+\rightarrow  2_2^+)$&120.0& 111.9&149.1& 121.7&162.9&108.1&187.7&110.9&240.3&  99.3\\

    \hline
  \end{tabular}
  \caption{B(E2) values (in arbitrary units) obtained, 
    for $N=50$, for the fitted IBM
    Hamiltonians (see text) compared with those provided by the
    different X(5)-models.}
  \label{tab-be2-comp}
\end{table}

\section{The critical Hamiltonian}
\label{sec-crit}
One of the most attractive features of the X(5)-models treated in
this work is that they are supposed to describe, at different
approximation levels, the critical point in the transition from
spherical to rigid axially deformed shapes. Since they are connected to a
given IBM Hamiltonian, as shown in the preceding section, this should
correspond to the critical point in the transition from spherical to
axially deformed  shapes, {\it i.e.} this Hamiltonian should produce an
energy surface with degenerated spherical and deformed minima.
Is this the case for the
fitted IBM Hamiltonians obtained in the preceding section?  Before
starting with the discussion it is necessary to establish a measure on
how close is a given IBM Hamiltonian to the critical point.

An energy surface can be associated to a given IBM Hamiltonian by
using the intrinsic state formalism \cite{Gino80b,Diep80a,Diep80b}
which introduces the shape variables $(\beta,\gamma)$ in the IBM. To
define the intrinsic state one has to consider that the dynamical
behavior of the system can be approximately described in terms of
independent bosons moving in an average field \cite{Duke84}. The
ground state of the system is written as a condensate, $|c\rangle$, of
bosons that occupy the lowest-energy phonon state, $\Gamma_c^\dag$:
\begin{equation}
  \label{GS}
  | c \rangle = \frac{1}{\sqrt{N!}} (\Gamma^\dagger_c)^N | 0 \rangle,
\end{equation}
where
\begin{equation}
  \label{bc}
  \Gamma^\dagger_c = \frac{1}{\sqrt{1+\beta^2}}~ \left (s^\dagger + \beta
    \cos     \gamma          \,d^\dagger_0          + \frac{1}{\sqrt{2}}~\beta
    \sin\gamma\,(d^\dagger_2+d^\dagger_{-2}) \right) .
\end{equation}
$\beta$ and $\gamma$ are variational parameters related with the shape
variables in the geometrical collective model \cite{Gino80b}.  The
expectation value of the Hamiltonian (\ref{ham1}) in the intrinsic
state (\ref{GS}) provides the energy surface of the system,
$E(N,\beta,\gamma)=\langle c|\hat H| c \rangle$.  This energy surface
in terms of the parameters of the Hamiltonian (\ref{ham1}) and the
shape variables can be readily obtained \cite{Isac81} (note that we
keep the $\kappa_4$ variable for completeness),
\begin{eqnarray}
  \label{Ener1}
  \langle c|\hat H | c \rangle&=&
  {\displaystyle {\frac{N\beta^2}{(1+\beta^2 )}}}
  \Bigl(\varepsilon_d+
  % K N +
  6 \,\kappa_1 
  -\frac{9}{4}\,\kappa_2+\frac{7}{5}\,\kappa_3+\frac{9}{5}
  \,\kappa_4\Bigr)\nonumber\\
  &+&{\displaystyle {\frac{N(N-1)}{{{(1+\beta^2)}^2}}}}
  \Big[\frac{\kappa_0}{4}+\beta^2(- \frac{\kappa_0}{2}+4\,\kappa_2)
  +2\,{\sqrt{2}}\,\beta^3\,\kappa_2\,\cos(3\,\gamma)
  \nonumber\\
  \qquad\qquad\qquad
  &&+\beta^4(\frac{\kappa_0}{4}
  +\frac{\kappa_2}{2}+\frac{18}{35}\,\kappa_4)\Big].
\end{eqnarray}

The shape of the nucleus is defined through the equilibrium value of
the deformation parameters, $\beta$ and $\gamma$, which are obtained 
minimizing the ground state energy, $\langle c|\hat{H}|c \rangle$. A
spherical nucleus has a global minimum in the energy surface at $\beta=0$,
while a deformed one has the absolute minimum at a finite value of
$\beta$. The parameter $\gamma$ represents the departure from axial
symmetry, {\it i.e.} $\gamma=0$ and $\gamma=\pi/3$ stand for an
axially deformed nucleus, prolate and oblate respectively, while any
other value corresponds to a triaxial shape. An additional situation
appears when the energy surface is independent on $\gamma$ but
shows a minimum at a finite value of $\beta$, 
in this case the nucleus is $\gamma$-unstable. It
has to be noted that for a general IBM Hamiltonian including up to two
body terms, as the one considered in this work, triaxiality is 
forbidden.

With the tools described above one can study phase transitions in the
IBM \cite{Diep80a}. First, the parameters that define the Hamiltonian
are the control parameters and are usually chosen in such a way that
only one of them is a variable, while the rest remain constant. The
deformation parameters $\beta$ and $\gamma$ become the order
parameters, although in our case the order parameter is just $\beta$.
Roughly speaking, a phase transition appears when there exists an
abrupt change in the shape of the system when changing smoothly the
control parameter. The phase transitions can be classified according
to the Ehrenfest classification \cite{Stan71}.  First order phase
transitions appear when there exists a discontinuity in the first
derivative of the energy with respect to the control parameter. This
discontinuity appears when two degenerate minima exist in the energy
surface for two values of the order parameter $\beta$. Second order
phase transitions appear when the second derivative of the energy with
respect to the control parameter displays a discontinuity. This
happens when the energy surface presents a single minimum for
$\beta=0$ and the surface satisfies the condition ${\left(\frac{d^2
      E}{d\beta^2}\right)_{\beta=0}}=0$. In a modern
classification, second order phase transitions belongs to the high
order or continuous phase transitions \cite{Stan71}. The X(5)
situation was designed to describe first order phase transitions.

To determine whether a given Hamiltonian corresponds to a critical
point or not, the flatness or the existence of two degenerate minima
in the energy surface should be investigated.  For the case of one
parameter IBM Hamiltonian, {\it e.g.}~Consistent Q (CQF) Hamiltonians
\cite{Warn82}, it is simple to find an analytical expression for the
critical control parameter in the Hamiltonian. However, for a general
IBM Hamiltonian, as the one used in this work, it is necessary to
rewrite the energy surface in a 
special way, as proposed first by L\'opez-Moreno and Casta\~nos in
Ref.~\cite{Lope96}. There, the 
authors manage to write the energy surface of a general IBM
Hamiltonian in terms of two parameters. They made use of some
concepts from the Catastrophe Theory \cite{Gilm81} to define the two
essential parameters, ($r_1,r_2$) of the problem. In terms of these
two essential parameters they found
expressions for the locus, in the essential parameter space, that
gives a critical point at the origin in $\beta$, called bifurcation
set, and for the locus that gives rise to two degenerate minima,
called Maxwell set. The essential parameters $r_1$ and $r_2$ can be written as,
\begin{equation}
  r_1=\frac{a_3-u_0+\tilde\varepsilon/(N-1)}{2 a_1+
    \tilde\varepsilon/(N-1)-a_3}, 
  \label{r1}
\end{equation}

\begin{equation}
  r_2=-\frac{2 a_2}{2 a_1+\tilde\varepsilon/(N-1)-a_3},  
  \label{r2}
\end{equation}
where
\begin{eqnarray}
\tilde\varepsilon&=& \varepsilon_d+
%K N + 
6\,\kappa_1 
-{9\over 4}\,\kappa_2+{7\over 5}\,\kappa_3+{9\over 5}
\,\kappa_4\nonumber\\
a_1&=&{1\over 4}\,\kappa_0+{1\over 2}\,\kappa_2+{18\over 35}\,\kappa_4
\nonumber\\
a_2&=&2\,\sqrt{2} \,\kappa_2
\nonumber\\
a_3&=&-{1\over 2}\,\kappa_0+4\,\kappa_2
\nonumber\\
u_0&=&{\kappa_0\over2}.
\label{coeff}
\end{eqnarray}

Using the essential parameters, the energy surface can be written as:
\begin{eqnarray}
\nonumber
E^*(\beta,\gamma)&=&\frac{\langle c|\hat H | c \rangle
  -N(N-1)u_0/2}{N(N-1)(2 a_1+\tilde\varepsilon/(N-1)-a_3)} \\
&=&\frac{1}{1+\beta^2}(\beta^4+r_1 \beta^2(\beta^2+2)-r_2 \beta^3 cos 3\gamma).
\label{ener-catas}
\end{eqnarray}
Note that $E^*(\beta,\gamma)$ does not depend on the number of bosons, 
allowing to compare fairly energy surfaces corresponding to
different boson numbers.

\begin{figure}[hbt]
  \centering
  \includegraphics[width=10cm]{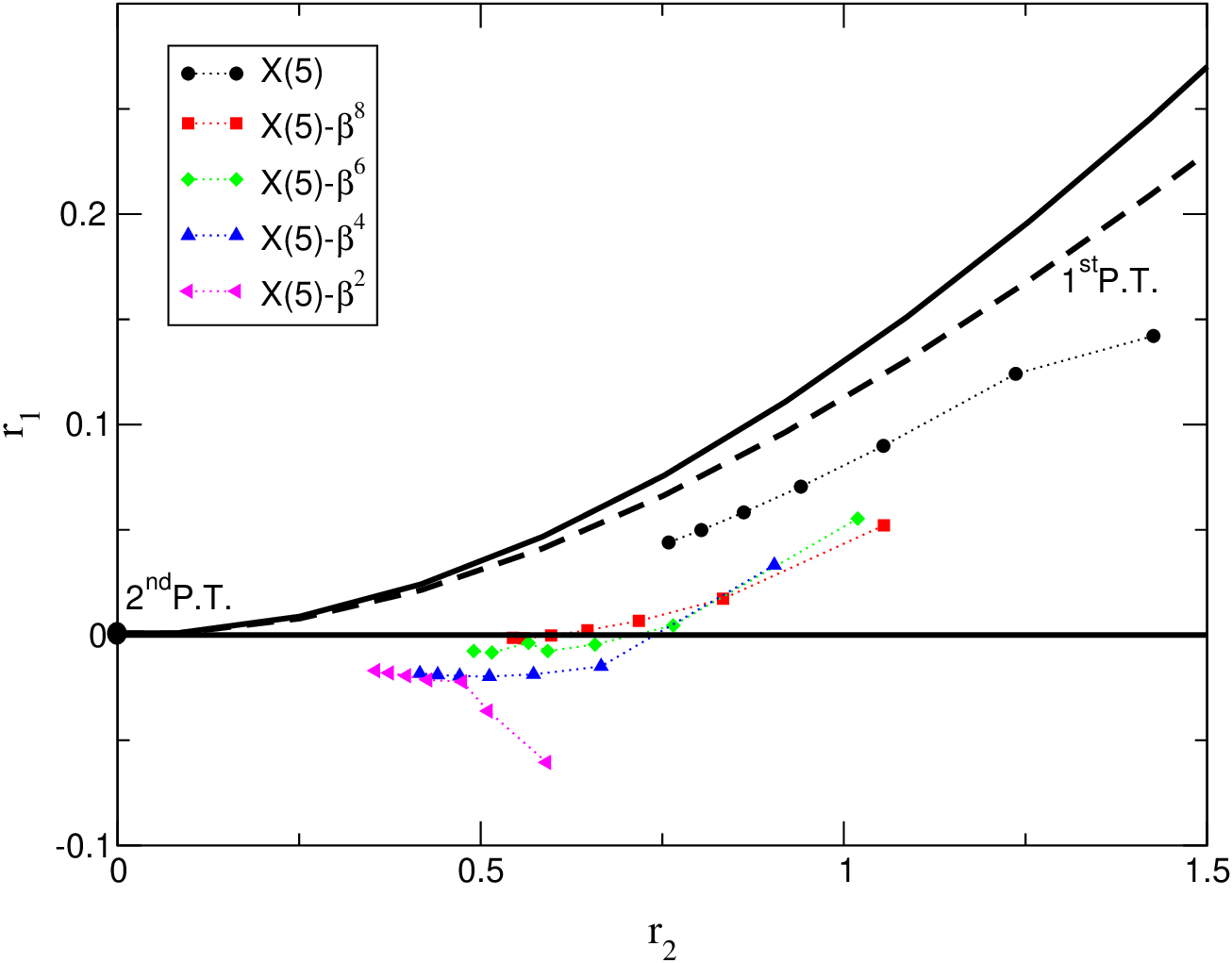}
  \caption{(Color online). Position in the plane $r_1-r_2$ of the
  energy surfaces extracted from the fitted IBM Hamiltonians to the
  different X(5) models  as a function of $N$ (10, 20, 30, 40, 50, 60,
  and 70). The largest value of N is placed on the left hand side
  whereas the smallest one is on the right part of the figure.}
  \label{fig-r1-r2}
\end{figure}
   
In figure \ref{fig-r1-r2} we display the plane of the essential 
parameters where the fitted Hamiltonians to the different X(5)-models
are plotted with different symbols.
In this plane a critical first order Hamiltonian corresponds to a point over
the dashed line (Maxwell set, {\it i.e.}~two degenerated minima). The
$r_1=0$ corresponds to $\left(\frac{d^2
    E}{d\beta^2}\right)_{\beta=0}=0$, {\it i.e.} to the appearance of
the spherical minimum (antispinodal point). The curved full line above the Maxwell set corresponds to
$\left(\frac{d^2
    E}{d\beta^2}\right)_{\beta_0}=0$, with $\beta_0\neq 0$, {\it i.e.} it
corresponds to the appearance of a deformed minimum (spinodal point). 
The area below
the curved full line and above $r_1=0$ is the coexistence region. In
this region two minima, one spherical and one deformed, coexist. 
We also represent in the plane 
%the three IBM dynamical symmetry limits and 
the second order critical point ($r_1=r_2=0$). 
Note that all the points above the dashed line correspond to
spherical shapes, while those below that line represent deformed shapes.
We
only plot the $r_2>0$ semi plane because our IBM Hamiltonian always
corresponds to prolate shapes and therefore to $r_2>0$. The semi plane
$r_2<0$ is identical to the presented in figure \ref{fig-r1-r2}, but
for oblate shapes. The different symbols in the figure correspond to the
IBM  Hamiltonians fitted to one of the X(5) models (see legend) for N
ranging from 10 to 70.  The idea is to see if in the 
large N limit the
obtained IBM Hamiltonians go to the Maxwell set (dashed line) that is the line of
first order phase transitions. For all the models, the N value of the different
points increases from the right to the left.

Several important features can be extracted from this figure. The main
one is that  the X(5)
case, for the whole range of $N$, is always very close to the
Maxwell set, {\it i.e.} to the first order phase transition line,
being closer as the value of $N$ increases. This is not exactly the case for
the rest of fits. 
X(5)-$\beta^8$, X(5)-$\beta^6$, and X(5)-$\beta^4$  seat
on the coexistence region for small or moderate N values, but they go towards
the prolate deformation region as $N$ increases, although they are
always very close to the antispinodal line. Finally, the case
X(5)-$\beta^2$ lies on the deformed region but it approaches to the
coexistence area as N increases. 
In general, one observes that the mapped IBM energy surfaces moves
further away from the coexistence region as one changes from X(5) through
X(5)-$\beta^8$, X(5)-$\beta^6$, X(5)-$\beta^4$, and  X(5)-$\beta^2$. 
In general, the higher the value of n is (X(5)-$\beta^n$), the closer
to the coexistence region is the IBM energy surface.
In all the cases one observes that the different models always move in
the direction of the second order phase transition point as N increases.

\begin{figure}[hbt]
  \centering
  \includegraphics[width=12cm]{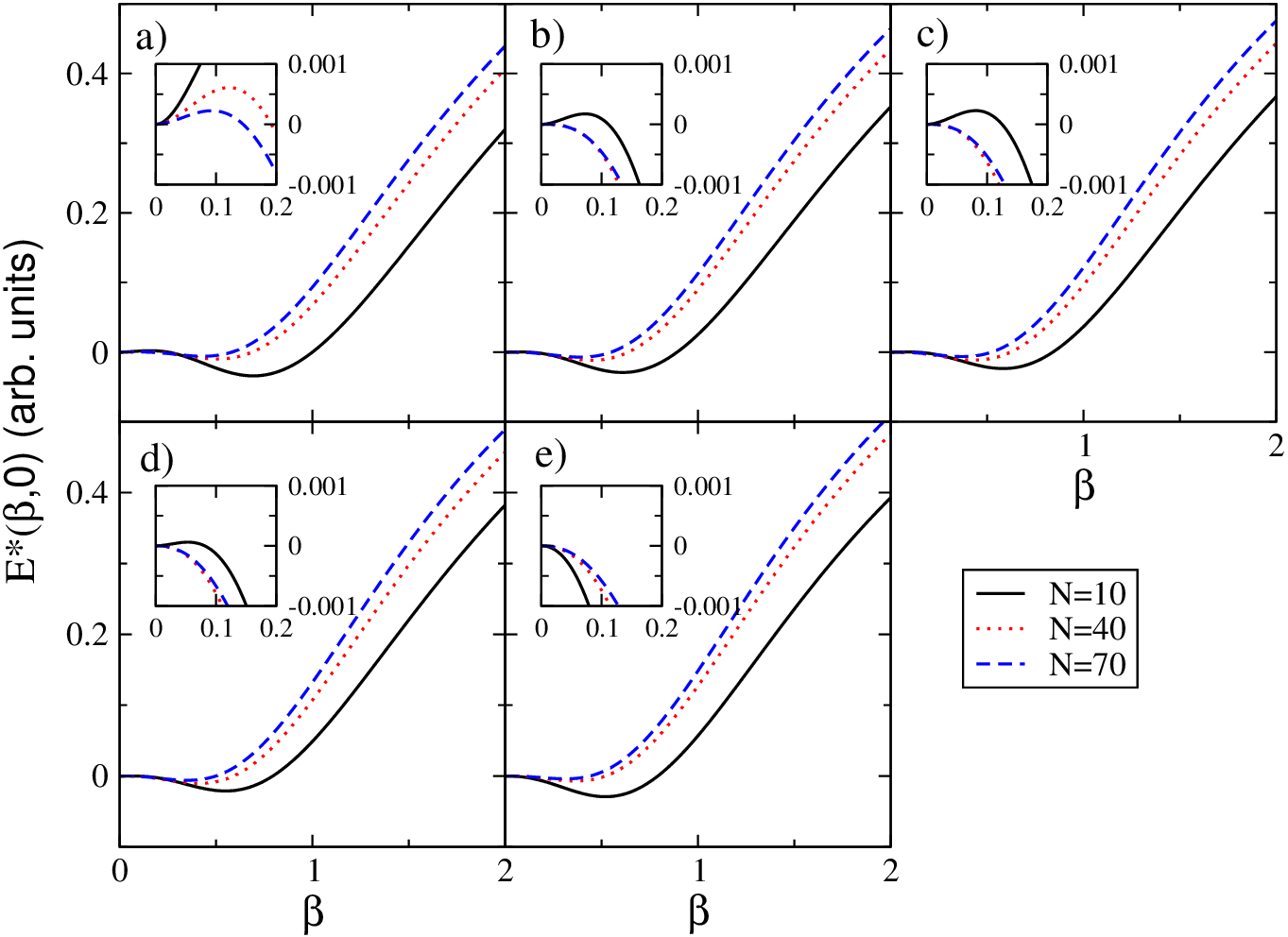}
  \caption{(Color online). IBM axial energy surfaces, for prolate
    shapes (see eq.~\ref{ener-catas}) as a function of
    $\beta$, for selected values of $N$ (see text for definition). a)
    X(5), b) X(5)-$\beta^8$, c) X(5)-$\beta^6$, d) X(5)-$\beta^4$, and
    e) X(5)-$\beta^2$. In the insets a closer view of the region
    around $\beta=0$ are shown.}
  \label{fig-ener-r1-r2}
\end{figure}

In order to illustrate graphically the shapes of the energy surfaces
obtained with the different IBM
Hamiltonians, we plot in figure \ref{fig-ener-r1-r2} the axial IBM energy
surfaces $E^*(\beta,0)$ (eq.~\ref{ener-catas}), along the prolate leg,
extracted from the fit to the five analyzed X(5)-models as a
function of the deformation $\beta$ for three values of N: 10, 40, and
70. All the panels show a rather similar aspect, for N=10 there exist
a more pronounced minimum, while passing from N=40 to N=70 the energy
surface flattens, moving into a shape with two approximately
degenerated minima. As it was studied in figure \ref{fig-r1-r2}, in
panel a), which is for X(5), the three curves present two minima, in
panels b), c), and d), corresponding to X(5)-$\beta^8$,
X(5)-$\beta^6$, and X(5)-$\beta^4$, respectively, only the line for
N=10  has two minima, while N=40 and N=70 show only a deformed
minimum (see insets of figure \ref{fig-ener-r1-r2}). 
Finally, panel e) is for X(5)-$\beta^2$, and the three curves
present only a deformed minimum, although it clearly flattens as N
increases. This analysis confirms that X(5)-models map into IBM
Hamiltonians that are very close to the first order
phase transition line, but the closest Hamiltonian to the transition area is
the one mapped from the original X(5) model.

A somehow similar analysis to the one presented in this section was
performed in \cite{Mccu05,Mccu06}. There, the authors studied the
connection between X(5) and X(5)-$\beta^2$ using a two parameter IBM
Hamiltonian (they also analyzed X(3)) for $N=10$. 
Their conclusion was that the X(5) case is close to, although not exactly
at, the first order phase transition region for a finite number of
boson. This conclusion is very similar to the one extracted from our results. For the case of
X(5)-$\beta^2$ their results are similar to the X(5) case, {\it i.e.}~the corresponding
IBM Hamiltonian is even closer to the first order phase transition
line that in the X(5) case. These later results 
are also in agreement with the conclusions raised in the present
work, although we have found that X(5)-$\beta^2$ is further from the
coexistence region than X(5).
The origin of the discrepancy should be based on the more
restricted set of data used in \cite{Mccu05,Mccu06} in their fits and
the smaller number of parameters used in the IBM Hamiltonian. 

\section{Summary and conclusions}
\label{sec-conclu}
In this paper, we have studied the connection between the
X(5)-models and the IBM on the basis of 
a numerical mapping between these 
models. To establish the mapping we have performed a best fit of the
general IBM Hamiltonian to a selected set of
energy levels produced by several X(5)-models; as an additional
parameter we have used the energy of the gamma band head, which is not fixed
by the X(5)-models.  Later on, a free parameter check of
the wave functions, obtained with the best fit parameters, has been
done by calculating relevant B(E2) transition rates.
All calculations have been done as a function of the number of bosons.
Once the best fit IBM Hamiltonians to the different X(5)-models are
obtained, their energy surfaces are constructed and analyzed with the
help of the Catastrophe Theory so as to know how close they are to a
critical point. 

We have shown that it is possible, in all cases, to obtain a
mapping between the X(5)-models and the IBM with a
reasonable agreement for both energies and B(E2) transition
rates.  In general, the goodness of the fit to the energies and B(E2)
transition rates is
independent on the number of bosons. 
Globally, the agreement is similar for all the models: for the
energies the best agreement is for X(5)-$\beta^8$ and
  X(5)-$\beta^6$ while the worst is for 
X(5). Anyway, the $\chi^2$ values obtained that give the goodness of
the fits are comparable for all the models. Additional tests have been
done to the produced wave functions by calculating B(E2) transition
rates. No free parameters are included in these calculations.
In this case, the best agreement (smallest  $\chi^2$ value) is obtained
for X(5) while the worst is for X(5)-$\beta^2$.
A consequence of this good general
agreement is that it would be almost impossible, 
from an experimental point of view,
to discriminate between a X(5)-model and its corresponding IBM
Hamiltonian when only few low-lying states are considered (usually the
four lowest states in the ground, beta, and gamma bands.) 

We have also proved that the X(5) model corresponds to an IBM
Hamiltonian which is very close to the first order phase transition region,
getting closer for larger values of $N$. X(5)-$\beta^8$,
X(5)-$\beta^6$, and X(5)-$\beta^4$  models seat on the coexistence
region for small and moderate values of
$N$, but they slightly move into the deformed region for the larger values of
$N$. 
Finally X(5)-$\beta^2$
stands all the way in the deformed region, although very close to
the coexistence region. In all the cases the system evolves towards the
second order critical point as N increases.

It is worth mentioning that the conclusions raised in this work are
somehow dependent on the constrains imposed in the fitting
procedure. On one hand we have checked that the inclusion, or not, of high lying
states in the bands considered in the fit does
not strongly affect to the results. On the other hand we have extensively
checked the case with $\kappa_4\neq 0$ \cite{Unpu10}. This later case
generates values of  
$\chi^2$ similar to the ones presented in this work, but the global picture
of the mapped IBM Hamiltonians to the X(5)-models is not so consistent
as one the shown here. In
particular, the change of the Hamiltonian parameters as a function of
N or as a function of the considered X(5)-model is not smooth
enough. There are instabilities in  the fitting procedure due to
the fact that the produced $\chi^2$ surface is very flat. 

Finally, it is worth mentioning the differences between
the X(5)-models/IBM and  the E(5)-models/IBM mapping \cite{Garc08}. 
For the E(5)-models the agreement for both, the energies and 
the B(E2) transition
rates is really remarkable and much better than for the X(5)-models.  
Globally, the best agreement is obtained for the
E(5)-$\beta^4$ Hamiltonian and the worst for the E(5) case.
For the case of very large number of bosons the only E(5)-model 
that can be reproduced exactly by the
IBM is E(5)-$\beta^4$, corresponding such a Hamiltonian with the
critical point of the model ($r_1=0$) (as shown in \cite{Aria03,Garc05}). 
All the E(5)-models correspond to IBM
Hamiltonians very close to the critical area, $|r_1|<0.05$ with $r_2=0$.
Therefore, one can say that the E(5)-models are appropriate to
describe transitional $\gamma-$unstable regions close to the critical
point. However, not all the X(5)-models are suitable for describing
the critical area between the axially deformed and the spherical
shapes, only X(5) is really appropriated to this end. 
Finally, for the E(5)-models it is observed the existence of
something similar to a quasidynamical symmetry \cite{Rowe04a},
we call this phenomenon quasi-critical point symmetry. In the case of
the X(5)-models we cannot talk about quasi-critical point symmetry
because the agreement between IBM and the X(5)-models is not good
enough, only in the case of the ground state band for X(5) we have the
appropriated agreement to say that a quasi-critical point symmetry is present.
  
\section{Acknowledgements}
This work has been partially supported
by the Spanish Ministerio de Educaci\'on y Ciencia and by the European
regional development fund (FEDER) under projects number FIS2008-04189,
FPA2007-63074, by CPAN-Ingenio and by the Junta de
Andaluc\'{\i}a under projects FQM160, FQM318, P05-FQM437 and
P07-FQM-02962.

\end{document}